\documentclass[10pt]{article}

\usepackage{graphicx}
\usepackage{caption} 
\usepackage{cite}
\usepackage[utf8]{inputenc}
\usepackage[T1]{fontenc}
\usepackage{authblk}
\usepackage{hyperref} 
\usepackage{xcolor} 
\usepackage[left=25mm,top=25mm,bottom=25mm,right=25mm]{geometry}
\title{ {\bf 
On new regular charged black hole solutions: Limiting Curvature Condition, Quasinormal modes and Shadows
} }

\author[1]{Leonardo Balart\thanks{leonardo.balart@ufrontera.cl}}

\author[1]{Grigoris Panotopoulos\thanks{grigorios.panotopoulos@ufrontera.cl}}

\author[2,3]{\'Angel Rinc\'on\thanks{angel.rincon@ua.es}}

\affil[1]{Departamento de Ciencias F{\'i}sicas, Universidad de La Frontera, Casilla 54-D, 4811186 Temuco, Chile.} 
\affil[2]{Departamento de Física Aplicada, Universidad de Alicante,
Campus de San Vicente del Raspeig, E-03690 Alicante, Spain.}
\affil[3]{Sede Esmeralda, Universidad de Tarapacá, Avda. Luis Emilio Recabarren 2477, Iquique, Chile}

\begin{document}

\date{}

\maketitle

\abstract{
We introduce two new static, spherically symmetric regular black hole solutions that can be obtained from non-linear electrodynamics models. For each solution, we investigate the dynamic stability with respect to arbitrary linear fluctuations of the metric and electromagnetic field, and also examine the energy conditions that those black holes satisfy. Moreover, based on those solutions, we present two additional ones that satisfy the Limiting Curvature Condition. 
Finally, we make a comparison between the two solutions exploring their null geodesics and circular photon orbits.
}

\section{Introduction}

Solutions of static regular black holes (BHs) are characterized by not presenting divergences in the metric function or in the curvature invariants~\cite{Bronnikov:2022ofk, Lan:2023cvz}. Various solutions have been built, for example Refs.~\cite{Bardeen:1968, Dymnikova:1992ux, AyonBeato:1998ub, Ayon-Beato:1999qin, AyonBeato:1999rg, Bronnikov:2000vy, Dymnikova:2004zc, Ayon-Beato:2004ywd, Hayward:2005gi, Nicolini:2005vd, Culetu:2013fsa, Balart:2014jia, Balart:2014cga, Ma:2015gpa, Frolov:2016pav, Fan:2016hvf, Rodrigues:2017yry, Rodrigues:2020pem, Kruglov:2021mfy}. Some of these models arise from the coupling of gravity with a nonlinear electrodynamic model and in several of them the Reissner-Nordstr{\"o}m solution can be recovered at large distances. 
The properties and characteristics of this type of static black hole solutions have been investigated in several studies. For example, see Refs.~\cite{Kruglov:2017fck, Contreras:2017eza, Ghosh:2018bxg, Mazharimousavi:2019jja, Junior:2020zdt, Melgarejo:2020mso, Singh:2020rnm, Kumar:2020bqf, Cai:2021ele, Singh:2021nvm, Singh:2022xgi, Kruglov:2021yya, Asl:2021krj, Kumar:2023cmo, dePaula:2023ozi, Balart:2023odm, Konoplya:2023aph, Konoplya:2023ahd, Guo:2023pob, Ovgun:2024zmt}.

\smallskip 

Different restrictions have been considered in the study of regular black holes for them to be deemed plausible~\cite{Bronnikov:2000vy, Moreno:2002gg, Burinskii:2002pz, Zaslavskii:2010qz, Frolov:2016pav, Maeda:2021jdc, Bokulic:2022cyk, Bronnikov:2022ofk, Bokulic:2023afx, Russo:2024xnh}. Some of them refer to conditions that the energy-momentum tensor of the Einstein field equations $T_{\mu\nu}$ must satisfy, which are known as energy conditions~\cite{Hawking:1973uf}. For the particular case of regular black hole solutions see, for example, Refs.~\cite{Dymnikova:2004zc, Maeda:2021jdc, Lan:2022bld}, where the following are considered standard conditions:
the null energy condition (NEC), the weak energy condition (WEC), the dominant energy condition (DEC), the strong energy condition (SEC).
Here, if we consider that $\xi_\mu$ and $k_\mu$ are arbitrary and null time vectors, respectively,
then the conditions mentioned for $T_{\mu\nu}$ are given by the following relations: $T^{\mu\nu} k_\mu k_\nu \geq 0 \,\,\,\mbox{(NEC)}$; $T^{\mu\nu} \xi_\mu \xi_\nu \geq 0     \,\,\,\mbox{(WEC)}$; $T^{\mu\nu} \xi_\mu \xi_\nu \geq 0   \mbox{ and } T^{\mu\nu} \xi_\mu \mbox{ is a non-spacelike vector} \,\,\, \mbox{(DEC)}$; and $T^{\mu\nu}\xi_\mu \xi_\nu - \frac{1}{2} T^{\mu}_{\,\,\,\mu} \xi^\nu \xi_\nu\geq 0 \,\,\,\mbox{(SEC)}$. 

\smallskip 

Another requirement focuses specifically on regular solutions obtained with nonlinear electrodynamics. In Ref.~\cite{Moreno:2002gg}, specific conditions are established to ensure the dynamic stability of the black hole solutions with respect to linear fluctuations of the metric and the electromagnetic field. However, it could turn out that an unstable solution becomes stable after a possible emission of gravitational and electromagnetic radiation~\cite{Moreno:2002gg}. We will postulate that the electric or magnetic charge cannot exceed a value without the black hole becoming unstable.

\smallskip 

The limiting curvature condition (LCC) was proposed as a constraint on the maximum values that the curvature invariants can reach in Refs.~\cite{Markov:1982, Markov:1984ii, Polchinski:1989ae} and has been explored in various studies on nonsingular black hole models~\cite{Frolov:2014wja, Frolov:2016pav, Frolov:2017rjz, Frolov:2017dwy, Colleaux:2017ibe, Bohmer:2019vff, Ali:2019bcn, Kruglov:2019wjv, Frolov:2021kcv, Frolov:2021vbg, Maeda:2021jdc, Kruglov:2021mfy, Frolov:2021afd, Frolov:2022fsl, Boos:2023icv}. The main motivation behind of these studies is to address the problem that the curvature of spacetime could exhibit unlimited growth as parameters like mass or charge increase without limit. The LCC establishes that there is a fundamental length scale that imposes constraints on the curvature invariants so that they remain bounded. That is, the curvature invariants are restricted by a universal value for all values of the solution parameters. This universal value is denoted by $|R| \leq \mathcal{B}\ell^{-2}$, where $R$ represents a curvature invariant, $\ell$ is a fundamental length scale, and $\mathcal{B}$ is a dimensionless constant that may depend on the type of curvature invariant but remains independent of specific solutions within the theory.

\smallskip 

An example of black hole that satisfies the LCC is the regular spherically symmetric Hayward solution whose metric function is given by~\cite{Hayward:2005gi} (see the analysis in Ref.\cite{Frolov:2016pav} or~\cite{Maeda:2021jdc})
\begin{equation}
f(r) = 1- \frac{2 M r^2 }{r^3 + 2 M \ell^2}
\, ,
\label{Hayward}
\end{equation}
where $M$ is the mass of the black hole and $\ell$ a fundamental length scale. 
In this case the absolute values of the curvature invariants comply with the restriction of not exceeding some finite value, even when $M \rightarrow \infty$. Note, for instance, that in this limit the Ricci scalar of the Hayward model converges to the value $12/\ell^2$. 

\smallskip 

Conversely, there are solutions of regular black holes that do not obey the LCC. In Ref.~\cite{Maeda:2021jdc} three types of spacetime are presented that are in this category. These solutions are based on three well-known solutions of regular charged black holes reported in Refs.~\cite{Bardeen:1968}, \cite{Dymnikova:2004zc} and~\cite{Fan:2016hvf}, respectively, wherein the replacement of the electric or magnetic charge by the parameter $\ell$ is carried out. 
In particular for one of them, the Bardeen type solution, with mass $M$ and parameter $\ell$ the Ricci scalar at $r = 0$ is $24 M/\ell^3$, which diverges as $M \rightarrow \infty$.

\smallskip 

In the same Ref.~\cite{Maeda:2021jdc}, the author proposes some conditions that regular black hole models should meet. One of them is the LCC. Two others refer to the standard energy conditions that the corresponding energy-momentum tensor should satisfy: i) in asymptotically flat regions; ii) on the event horizon of a large black hole.

\smallskip 

Ideal BHs are supposed to be isolated objects. Realistic BHs of Nature, however, are in constant interaction with their environment. We may think, for instance, matter accretion onto a BH from its donor in binaries. When a black hole is perturbed due to a certain interaction, the geometry of space-time undergoes damped oscillations. How a system responds to small perturbations as well as normal modes of oscillating systems have always been important topics in physics. Regarding BH physics in particular, the work of \cite{regge} long time ago marked the birth of black hole perturbation theory, and later on it was extended by other people \cite{zerilli1,zerilli2,zerilli3,moncrief,teukolsky}. Nowadays the state-of-the art in black hole physics and perturbations is nicely summarized in the comprehensive review of Chandrasekhar's monograph \cite{monograph}. The information on how a given BH relaxes after the perturbation has been applied is encoded into the quasi-normal (QN) frequencies. The latter are complex numbers, with a non-vanishing imaginary part, that depend on the details of the background geometry as well as the spin of the propagating field at hand (scalar, Dirac, vector (electromagnetic), tensor (gravitational)), but they do not depend on the initial conditions. Therefore, QN modes (QNMs) carry unique information about black hole physics. Black hole perturbation theory and QNMs of black holes are relevant during the ringdown phase of binaries, in which after the merging of two black holes a new, distorted object is formed, while at the same time the geometry of space-time undergoes damped oscillations due to the emission of gravitational waves.

\smallskip 

From the experimental and observational point of view, a number of advances over the last 10 years or so have led to the direct observation of black holes. To be more precise, several years ago the LIGO collaboration directly detected for the first time the gravitational waves emitted from a BH merger of $\sim 60~M_{\odot}$ \cite{ligo}. However, at the time there was no information on the BH horizon, which is the defining property of BHs. After the LIGO historical direct detection, a few years ago the Event Horizon Telescope (EHT) project \cite{project} observed a characteristic shadow-like image \cite{L1}, see also \cite{L2,L3,L4,L5,L6} for physical origin of the shadow, data processing and calibration, instrumentation etc. That image was a darker region over a brighter background, via strong gravitational lensing and photon capture at the horizon. Thus, the BH shadow and its observation allows us to probe the space-time geometry in the vicinity of the horizon, and doing so we may test both the existence and the properties of the latter \cite{psaltis}. One should bear in mind, however, that other horizon-less objects that possess light rings also cast shadows \cite{horizonless1,horizonless2,horizonless3,horizonless4,horizonless5,horizonless6,shakih2018,shakih2019},
and therefore the presence of a shadow does not necessarily implies that the object is indeed a BH. Therefore, shadows as well as strong lensing \cite{vib1,vib2} images provide us with the exciting possibility a) to detect the nature of a compact object, and b) to test whether or not the gravitational field around a compact object is described by a rotating geometry. For a brief review on shadows see \cite{review}. The shadow of the Schwarzschild geometry was considered in \cite{synge,luminet}, while the shadow cast by the Kerr solution was studied in \cite{bardeen} (see also \cite{monograph}). For shadows of Kerr BHs with scalar hair see \cite{carlos1,carlos2}, and for BH shadows in other frameworks see 
\cite{Bambi:2008jg,Bambi:2010hf,study1,study2,Moffat,quint2,study3,Schee:2016yzb,study4,study5,Schee:2017hof,study6,bobir2017,kumar2018,ovgun2018,Konoplya:2019sns,sudipta2019,shakih2019b,contreras2019rot,sabir19,sunnya,sunnyb,Konoplya:2019xmn,Konoplya:2019fpy,Allahyari:2019jqz,Tinchev:2019qwt,Cunha:2019ikd,Ovgun:2019jdo,Konoplya:2019goy,Hensh:2019ipu,Stuchlik:2019uvf,Contreras:2019cmf,Fathi:2019jid,Chang:2020lmg,Badia:2020pnh,Li:2020drn,Li:2019lsm,Khodadi:2020jij,Ovgun:2020gjz,Liu:2020ola,Vagnozzi:2020quf,Ghosh:2020ece,Rincon:2023hvd,Ovgun:2023ego,Contreras:2020kgy}.
Finally, see~\cite{KumarWalia:2022ddq} for observational predictions and constraints
from supermassive black holes of polymerized regular black holes in the context of non-
linear electrodynamics, where the author has carried out a comprehensive study of static,
spherically symmetric polymerized black holes motivated by the Loop Quantum Gravity
principles and semi-polymerization technique.

\smallskip 

In this paper, we introduce two solutions of charged regular black holes. We analyze their stability. In particular, we determine the range of parameters that allows its stability. We also investigate the energy conditions satisfied by the energy-momentum tensor of each model, studying the respective parameter ranges in which they satisfy the null energy condition (NEC), the weak energy condition (WEC) and the dominant energy condition (DEC). Following Ref.~\cite{Maeda:2021jdc}, we take advantage these two charged black hole solutions to construct two models that obey the LCC, by substituting the electric charge with a length scale parameter.
Let us mention that, unlike the other known models, one of these solutions, as will be shown, allows in principle an extremal case whose value can exceed twice the mass of the black hole. This same solution, unlike other regular charged black hole, presents instability for some values of electric charge from the point of view of dynamic stability with respect to arbitrary linear fluctuations of the metric and electromagnetic field. Additionally, the two solutions presented, unlike most other known solutions (see~\cite{Maeda:2021jdc}), maintain regularity in their curvature invariants even when $M \rightarrow \infty$. Let us also point out that we have chosen form for each metric function $f(r)$ so that they not only exhibit regularity but are also analytically manageable and allow the fulfillment of some energy conditions.

\smallskip 

It should be noted that when considering electric and magnetic fields as sources of the same metrics, it is desirable to emphasise that the solutions then belong to different versions of NED. Moreover, the electric versions require different $L(F)$, hence different NED theories near the regular centre and at large radii, and suffer from other shortcomings,
as described in ref. [8].

\smallskip

In the present article we organize our work as follows: After this introductory section, we present the regular charged black hole solutions in section 2, while in the third section, we present the solutions that respect the Limiting Curvature Condition. In section 4 we briefly discuss null geodesics and circular photon orbits as well the QN spectrum in the eikonal limit adopting the WKB approximation. Finally, we conclude our work in the fifth section. Throughout the manuscript, we set the universal constants to unity, $G=1=c$, and we consider the mostly positive metric signature in four-dimensional space-time $\{ -,+,+,+ \}$.

\section{Regular charged black hole solutions}

\subsection{First solution}

We will analyze some characteristics of two regular charged black solutions. In both cases, we consider the following stationary and spherically symmetric metric
\begin{equation}
ds^2 = -f_i(r) dt^2 +  f_i^{-1}(r) dr^2 + r^2 (d\theta^2 + \sin^2 \theta \, d\phi^2)
\, .
\label{metric}
\end{equation}
If $m_i(r)$ is the mass function, then we can write
\begin{equation}
f_i(r) = 1 - \frac{2 m_i(r)}{r}
\, .
\label{metric}
\end{equation}

The first metric function that we present is given by the following expression
\begin{equation}
f_1(r) =  1-\frac{432 M^4 r^2}{432 M^4 q^2+\left(6 M r+q^2\right)^3}
\, ,
\label{1-sol}
\end{equation}
This solution asymptotically behaves as 
\begin{equation}
f_1(r) \sim  1 -\frac{2 M}{r}+\frac{q^2}{r^2} -\frac{q^4}{3 M r^3}
\, ,
\label{inf-1-sol}
\end{equation}
and in the limit $r \rightarrow 0$, we have that
\begin{equation}
f_1(r) \sim 1 - \frac{432  M^4 r^2}{q^2 \left(432 M^4 + q^4\right)}
\, ,
\label{zer-1-sol}
\end{equation}

\begin{figure}[h!]
\centering
\includegraphics[scale=0.8]{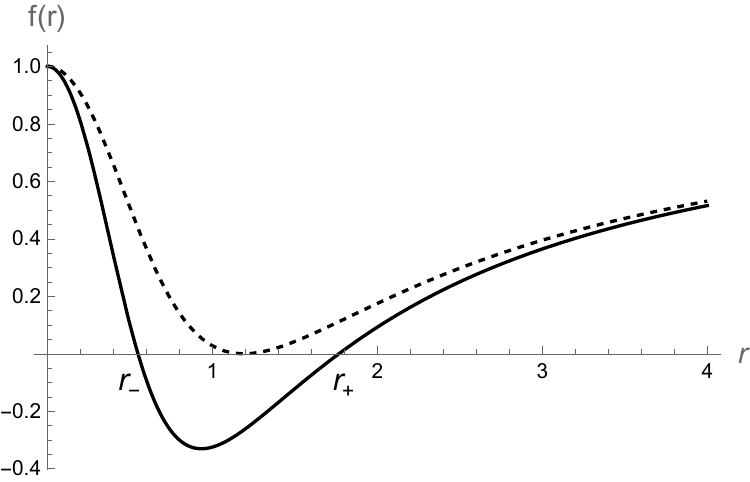}
\caption{The solid line illustrates the metric function for a case where $q < 0.6458 M$, here $r_+$ is the event horizon and $r_-$ the Cauchy horizon. The dashed line represents the extreme black hole. In this figure we choose $M = 1$, $q = 0.45$ (solid line) and $q = q_{ext} = 0.6458$ (dashed line).}
\label{metric-f1}
\end{figure}

This metric function has two roots when $q < q_{ext} = 0.6458 M$. The extreme black hole occurs when $q = q_{ext}$. In Fig.~(\ref{metric-f1}) we sketch these two cases.

If we consider $q$ as a magnetic charge then the Lagrangian corresponding to the model given in Eq.~(\ref{1-sol}) is
\begin{equation}
\mathcal{L}(\mathcal{F}) = -\frac{81 \mathcal{F} \left[  \left(s^4+27\right) \sqrt{2 q^2 \mathcal{F}}  + 6\ s^3 (2 q^2 \mathcal{F})^{1/4}  + 9  s^2\right]}
{\left[\left(s^4 + 27\right) (2 q^2 \mathcal{F})^{3/4} + 9\ s^3 \ \sqrt{2 q^2 \mathcal{F}} + 27  s^2 (2 q^2 \mathcal{F})^{1/4}  + 27 \ s\right]^2}
\, ,
\label{Lagr-1}
\end{equation}
where $s = q/(2M)$ and
\begin{equation}
\mathcal{F} = \frac{1}{4} F_{\mu\nu} F^{\mu\nu} = \frac{q^2}{2 r^4}
\, .
\label{F-1}
\end{equation}
In the limit of weak fields one obtains $\mathcal{L} \rightarrow \mathcal{F}$.
To obtain the corresponding Hamiltonian, one can perform a Legendre transformation~\cite{Salazar:1987}.

If we consider $q$ as an electric charge then the Hamiltonian is read as
\begin{equation}
\mathcal{H}(\mathcal{P}) = \frac{81 \mathcal{P} \left[  \left(s^4 + 27\right) \sqrt{-2 q^2 \mathcal{P}}  + 6\ s^3 (-2 q^2 \mathcal{P})^{1/4}  + 9  s^2\right]}
{\left[\left(s^4 + 27\right) (-2 q^2 \mathcal{P})^{3/4} + 9\ s^3 \ \sqrt{-2 q^2 \mathcal{P}} + 27  s^2 (-2 q^2 \mathcal{P})^{1/4}  + 27 \ s\right]^2}
\, .
\label{Hamil-1}
\end{equation}
Here
\begin{equation}
\mathcal{P} = \frac{1}{4} P_{\mu\nu} P^{\mu\nu} = - \frac{q^2}{2 r^4}
\, .
\label{P-1}
\end{equation}
In the limit of weak fields one obtains $\mathcal{H} \rightarrow \mathcal{P}$.

It should be noted that when we consider electric and magnetic fields as sources of the same metrics, the resulting solutions correspond to different models of nonlinear electrodynamics. Consequently, a Legendre transformation does not relate the Lagrangian given in Eq.~(\ref{Lagr-1}) to the Hamiltonian in Eq.~(\ref{Hamil-1}). Here, let us further add that, as pointed out in Ref.~\cite{Bronnikov:2000vy}, the electric solution versions present various problems; for example, they require different forms of $\mathcal{L}(\mathcal{F})$ in distinct regions to satisfy the regularity condition, unlike the magnetic solutions.


\subsubsection*{$\bullet$ Dynamic stability}

We can study the dynamic stability of this regular black hole solution with respects to arbitrary linear fluctuations of the metric and electromagnetic field, writing the Lagrangian~(\ref{Lagr-1}) in terms of the variable $x = q^2/r^2$, that is
\begin{equation}
\mathcal{L}(x) = \frac{81 x^2 \left[\left(s^4+27\right) x+6 s^3 \sqrt{x}+9 s^2\right]}{2 q^2 \left[\left(s^4+27\right) x^{3/2}+9 s^3 x+27 s^2 \sqrt{x}+27 s\right]^2}
\, ,
\label{Lagr-1-x}
\end{equation}
and verifying if the following inequalities are satisfied for all $0 < x <x_h$, where $x_h = q^2/r_h^2$~\cite{Moreno:2002gg}
\begin{eqnarray}
\mathcal{L} &>& 0 \ ,
\label{inq-1}
\\
\mathcal{L}_x &>& 0 \ ,
\label{inq-2}
\\
\mathcal{L}_{xx} &>& 0 \ ,
\label{inq-3}
\\
3 \mathcal{L}_x &\geq& x f(x) \mathcal{L}_{xx}   \ , .
\label{inq-4}
\end{eqnarray}
Here the first two inequalities are easily verified, while for the third and fourth their verification is analogous to what is shown in the graphs presented for the subsequent black hole solution.


\subsubsection*{$\bullet$ Curvature invariants}
In addition to the regularity of the metric, it can be shown that the curvature invariants are also regular~\cite{Rodrigues:2023fps}. Furthermore, upon conducting a more specific analysis, when computing the Ricci scalar $R$ and Kretschmann scalar $K$, we obtain
\begin{eqnarray}
R &=& -3456 M^4 \Bigg[5184 M^4 q^2 \Bigg(\left(432 M^4 q+q^5\right)^2+24 M q^4 r \left(432 M^4+q^4\right)
\nonumber \\
&&
+1296 M^4 q^2 r^4+864 M^3 r^3 \left(q^4-54 M^4\right)+216 M^2 q^2 r^2 \left(108 M^4+q^4\right)\Bigg)\Bigg]
\nonumber \\
&& \times
\left(432 M^4 q^2+216 M^3 r^3+108 M^2 q^2 r^2+18 M q^4 r+q^6\right)^{-3}
\, ,
\label{R-1}
\end{eqnarray}
\begin{eqnarray}
K &=& 4478976 M^8 \Bigg[1088391168 M^{12} r^{12}+1088391168 M^{11} q^2 r^{11}+574428672 M^{10} q^4 r^{10}+131010048 M^8 q^8 r^8
\nonumber \\
&&
+48 M q^{10} r \left(432 M^4+q^4\right)^3+q^8 \left(432 M^4+q^4\right)^4
+40310784 M^9 q^2 r^9 \left(7 q^4-216 M^4\right)
\nonumber \\
&&
+1944 M^4 q^8 r^4 \left(4292352 M^8+48384 M^4 q^4+107 q^8\right)+45349632 M^7 q^6 r^7 \left(96 M^4+q^4\right)
\nonumber \\
&&
+162 M^2 q^8 r^2 \left(432 M^4+q^4\right)^2 \left(288 M^4+7 q^4\right)+139968 M^6 q^4 r^6 \left(559872 M^8+16416 M^4 q^4+77 q^8\right)
\nonumber \\
&&
+93312 M^5 q^6 r^5 \left(419904 M^8+6264 M^4 q^4+19 q^8\right)
\nonumber \\
&&
+216 M^3 q^6 r^3 \left(432 M^4+q^4\right) \left(-186624 M^8+15984 M^4 q^4+83 q^8\right)\Bigg]
\nonumber \\
&& \times
\left(432 M^4 q^2+216 M^3 r^3+108 M^2 q^2 r^2+18 M q^4 r+q^6\right)^{-6}
\, .
\label{K-1}
\end{eqnarray}

One can observe that the Ricci scalar reaches its maximum value of $(5184 M^4)/(q^2 (432 M^4 + q^4))$ at the origin, and similarly, the Kretschmann scalar attains its maximum value of $(4478976 m^8)/(q^4 (432 m^4 + q^4)^2)$ at the same point.
Additionally, as $M$ approaches infinity, both $R$ and $K$ tend towards $12/q^2$ and $ 24/q^4$ respectively. Moreover, when we also consider the limit as $q \rightarrow \infty$, both $R$ and $K$ approach zero.


\subsubsection*{$\bullet$ Energy conditions}
Considering our gravitational source as a perfect fluid, we can express the energy conditions mentioned in the Introduction through the following inequalities: $\rho + p_i \geq 0$, $i=1,2,3$  (NEC), $\rho \geq 0$ and NEC (WEC), $\rho - p_i \geq 0$ and WEC (DEC), $\rho + p_1 + p_2 + p_3 \geq 0$ (SEC), where $\rho$ represents the density and $p_i$ denotes the pressures, which are given by 
\begin{equation}
\rho = - p_1 = T_{\,\,\,0}^0 = T_{\,\,\,1}^1 = \frac{m_i'(r)}{4 \pi r^2} \,\,\,\,\,\, , \,\,\,\,\,\,\,\, p_2 = p_3 = T_{\,\,\,2}^2 = T_{\,\,\,3}^3 = -\frac{m_i''(r)}{8 \pi r}
\, ,
\label{T-rho-p}
\end{equation}

From the above, we obtain that
\begin{equation}
\rho = \frac{1296 M^4 q^2 r \left(12 M \left(36 M^3+3 M r^2+q^2 r\right)+q^4\right)}{4 \pi\left(432 M^4 q^2+216 M^3 r^3+108 M^2 q^2 r^2+18 M q^4 r+q^6\right)^2} 
\, ,
\label{wec-s1-1}
\end{equation}
and
\begin{equation}
\rho + p_2 = \rho + p_3 = \frac{15552 M^5 q^2 r^2 \left(6 M r+q^2\right) \left(3888 M^5 r+432 M^4 q^2+216 M^3 r^3+108 M^2 q^2 r^2+18 M q^4 r+q^6\right)}{4 \pi \left(432 M^4 q^2+216 M^3 r^3+108 M^2 q^2 r^2+18 M q^4 r+q^6\right)^3} 
\, .
\label{wec-s1-2}
\end{equation}
Then it is straightforward to note that the first LCC solution satisfies the WEC for all $r$ and $M > 0$.

Given that the WEC is verified everywhere, to determine the region where the DEC is satisfied, we only need examine the behavior of the quantity $\rho-p_i$ with respect to the radial coordinate. This analysis yields the following result
\begin{eqnarray}
&&\rho - p_2 = \rho - p_3 =
\nonumber \\ &&
2592 M^4 q^2 r \Big[\left(432 M^4 q+q^5\right)^2+24 M q^4 r \left(432 M^4+q^4\right) + 1296 M^4 q^2 r^4 + 864 M^3 r^3 \left(q^4-54 M^4\right)
\nonumber \\
&&
 + 216 M^2 q^2 r^2 \left(108 M^4+q^4\right)\Big]
\times (4 \pi)^{-1}  \left(432 M^4 q^2+216 M^3 r^3+108 M^2 q^2 r^2+18 M q^4 r+q^6\right)^{-3} 
\, .
\label{dec-s1-3}
\end{eqnarray}
We immediately notice that the DEC is satisfied in the asymptotically flat region, since
\begin{equation}
\rho - p_2  = \rho - p_3 = \frac{q^4}{12 \pi M r^4 } +  O\left(r^{-6}\right)
\, ,
\label{}
\end{equation}
as $r \rightarrow \infty$.
When we do a numerical analysis of inequality~(\ref{dec-s1-3}), we  find that in the range $ 0.642245 M \geq q \geq q_{ext} = 0.6458 M$, the energy-momentum tensor satisfies the DEC on the event horizon.

\begin{figure}[h!]
\centering
\includegraphics[scale=0.9]{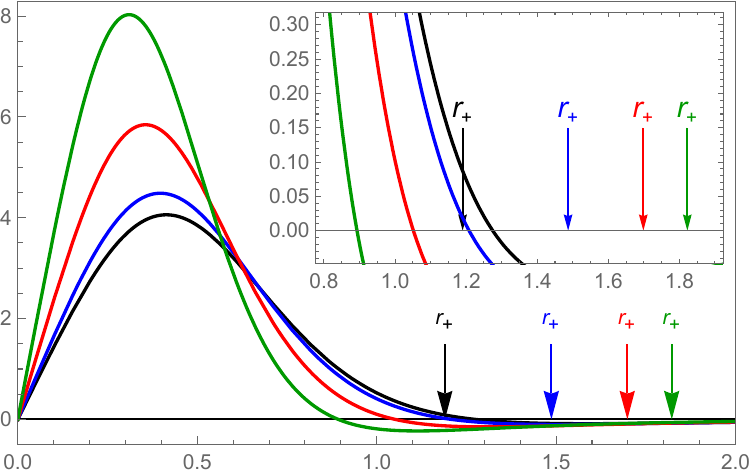}
\caption{A plot of the quantity $4\pi(\rho - p_2)$ as a function of $r$ is shown for various values of $q$ for the metric function $f_1(r)$. 
We choose $M = 1$ and $q = 0.4, 0.5, 0.6$ (green, blue, red) and $q = q_{ext} = 0.6458$ (black).
Colored vertical arrows indicate the corresponding event horizons. In the expanded view box it is observed that as the electric charge decreases, the condition depicted in the figure ceases to be met (it takes negative values) at the corresponding event horizon, unlike what occurs with the black curve.}
\label{DEC-f1}
\end{figure}

In Fig.~\ref{DEC-f1} we display different curves of $4\pi (\rho-p_2)$ as a function of the radial coordinate. Among the values of $q$ depicted in the figure, only the extreme black hole case exhibits an event horizon located within the region where $\rho - p_2 \geq 0$. Yet, this also occurs for values of the charge within the range $0.642245 \geq q \geq q_{ext} = 0.6458$, if we consider $M = 1$.

To examine the SEC compliance, we consider the relationships $\rho + p_i \geq 0$ and $\rho + p_1 + p_2 + p_3 \geq 0$. In particular for this second inequality, although it does not hold throughout the space as expected in advance, we can nevertheless find that for $M >0 $ and $ 0< q \leq q_{ext}$ the following inequality is satisfies for all $r$ from the event horizon to infinity
\begin{eqnarray}
\rho + p_1 + p_2 +p_3 &=& 1296 M^4 q^2 (4 \pi r)^{-1} [- \left(432 M^4 q+q^5\right)^2 - 18 M q^4 r \left(432 M^4+q^4\right) 
\nonumber \\
&& + 72 M^2 q^2 r^2 \left(216 M^4 - q^4\right)
+ 432 M^3 r^3 \left(432 M^4+q^4\right) + 3888 M^4 q^2 r^4 + 7776 M^5 r^5]
\nonumber \\
&&
\times \left(432 M^4 q^2+216 M^3 r^3+108 M^2 q^2 r^2 + 18 M q^4 r+q^6\right)^{-3} \geq 0
\, .
\label{}
\end{eqnarray}
Consequently, the SEC holds from the event horizon to infinity for all values of $q$ that allow the black hole solution to have an event horizon.

\subsection{Second solution}

The following metric function that we present is given by 
\begin{equation}
f_2(r) = 1-\frac{2 M}{r} \left(1 -\frac{M q^2}{M q^2+8 r^3}-\frac{q^2 r^3}{2 M\left(q^2+r^2\right)^2}\right)
\, ,
\label{2-sol}
\end{equation}
In $r \rightarrow \infty$ this solution behaves as 
\begin{equation}
f_2(r) \sim  1 - \frac{2 M}{r} + \frac{q^2}{r^2} + \frac{M^2 q^2-8 q^4}{4 r^4}
\, .
\label{inf-2-sol}
\end{equation}
In the limit $r \rightarrow 0$ the expression approaches 
\begin{equation}
f_2(r) \sim 1-\frac{15 r^2}{q^2}
\, .
\label{zer-2-sol}
\end{equation}
Solving the equation $f_2(r) = 0$ we obtain two horizons if $q < q_{ext} = 2.537862 M$.
The extremal black hole with degenerate horizon corresponds to the case where $q = q_{ext}$. 

Considering the same notation as above, if $q$ is a magnetic charge then the Lagrangian corresponding to this model is given by
\begin{equation}
\mathcal{L}(\mathcal{F}) = - \frac{3 \left(2 q^2 \mathcal{F}\right)^{3/2} - (2 q^2 \mathcal{F})}{2 q^2 \left(\sqrt{2 q^2 \mathcal{F}}+1\right)^3}
+ \frac{24 \left(2 q^2 \mathcal{F} \right)^{3/2}}{q^2 \left(\left(2 q^2 \mathcal{F}\right)^{3/4} + 16 s\right)^2}
\, ,
\label{Lagr-2}
\end{equation}
Note that in the limit of weak fields the Maxwell model $\mathcal{L} \rightarrow \mathcal{F}$ is recovered.

If we consider $q$ as electric charge then using the notation defined above we obtain the following Hamiltonian
\begin{equation}
\mathcal{H}(\mathcal{P}) = \frac{3 \left(-2 q^2 \mathcal{P}\right)^{3/2}- (-2 q^2 \mathcal{P})}{2 q^2 \left(\sqrt{-2 q^2 \mathcal{P}}+1\right)^3}-\frac{24 \left(-2 q^2 \mathcal{P} \right)^{3/2}}{q^2 \left(\left(-2 q^2 \mathcal{P}\right)^{3/4} + 16 s\right)^2}
\, .
\label{Hamil-2}
\end{equation}
In the limit of weak fields one obtains $\mathcal{H} \rightarrow \mathcal{P}$.

Again we define the variable $x = q^2/r^2$ and the value $x_+ = q^2/r_+^2$ to study the dynamic stability, that is, to study the fulfillment of the inequalities~(\ref{inq-1}-\ref{inq-4}), where
\begin{equation}
\mathcal{L}(x) = \frac{1}{2 q^2}\left(\frac{48 x^3}{\left(16 s+x^{3/2}\right)^2}+\frac{x^2-3 x^3}{(x+1)^3}\right)
\, .
\label{Lagr-2-x}
\end{equation}


\subsubsection*{$\bullet$ Dynamic stability}

\begin{figure}[h!]
\centering
\includegraphics[scale=0.6]{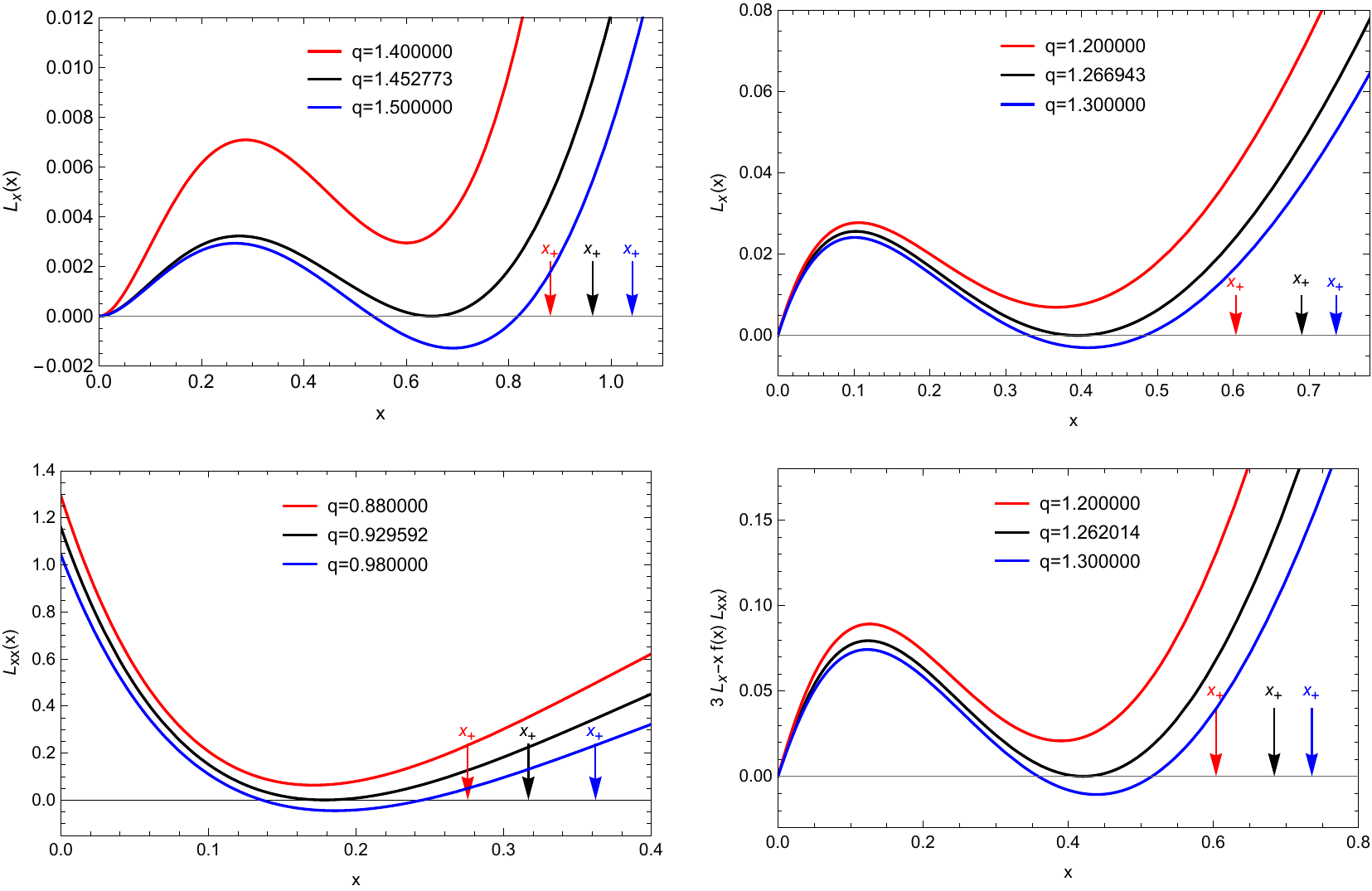}
\captionsetup{font={small,stretch=1.2}}
\caption{For each of the conditions indicated in Eqs.~(\ref{inq-1}) to~(\ref{inq-4}), the black curve, corresponding to $q = q_b$, represents the limiting case that separates the red curves, which satisfy each condition, from the blue curves, which do not. The points corresponding to $x_+$ on the horizontal axis are also shown in their respective colors. In the blue curves, where $q > q_b$, an additional interval within the range $[0, x_+]$ is observed where the conditions are not satisfied. The value of $q$ considered for each case is indicated, with $M = 1$.}
\label{Stability-S2}
\end{figure}

In our numerical analysis for $0 < x < x_h$ we find that the following inequalities are verified: $\mathcal{L} > 0$ if $0 \leq q < 1.452773 M$; $\mathcal{L}_x > 0$ if $0 \leq q < 1.2669433 M$; $\mathcal{L}_{xx} > 0$ if $0 \leq q < 0.929592 M$;
and $3 \mathcal{L}_x \geq x f(x)\mathcal{L}_{xx}$ if $0 \leq q < 1.262014 M$. This allows us to establish that this black hole is unstable if $0.929592 M \leq q \leq q_{ext}$.
Fig.~\ref{Stability-S2} displays the graphs in which we examine whether the inequalities given by Eqs.~(\ref{inq-1}) to~(\ref{inq-4}) are satisfied, respectively.

\subsubsection*{$\bullet$ Curvature invariants}

Additionally by computing the Ricci scalar $R$ and Kretschmann scalar $K$ we obtain respectively
\begin{eqnarray}
R &=& 
12 q^2 \left[q^2 \left(\frac{24 M^3}{\left(M q^2+8 r^3\right)^3}+\frac{1}{\left(q^2+r^2\right)^3}\right)-\frac{8 M^2}{\left(M q^2+8 r^3\right)^2}-\frac{2 q^4}{\left(q^2+r^2\right)^4}\right]
\, .
\label{R-2}
\end{eqnarray}

\begin{eqnarray}
K &=&  4 \left(\frac{4 \left(M^2 \left(15 q^8+49 q^6 r^2+48 q^4 r^4+16 q^2 r^6\right)-16 M \left(5 q^6 r^3+11 q^4 r^5+12 q^2 r^7+4 r^9\right)+64 q^2 r^6 \left(r^2-q^2\right)\right)^2}{\left(q^2+r^2\right)^6 \left(M q^2+8 r^3\right)^4}\right) \nonumber \\
&& + \frac{1}{\left(q^2+r^2\right)^8 \left(M q^2+8 r^3\right)^6} \times
\nonumber \\&&
\left[M^3 q^4 \left(15 q^8+72 q^6 r^2+93 q^4 r^4+64 q^2 r^6+16 r^8\right)-8 M^2 q^2 r^3 \left(115 q^8+424 q^6 r^2+681 q^4 r^4+448 q^2 r^6+112 r^8\right)\right.\nonumber \\
&& + \left. 64 M r^6 \left(13 q^8+88 q^6 r^2+87 q^4 r^4+64 q^2 r^6+16 r^8\right)-512 q^2 r^9 \left(q^4-8 q^2 r^2+3 r^4\right)\right]^2 \nonumber \\
&& + \left(\frac{q^2}{\left(q^2+r^2\right)^2}-\frac{16 M}{M q^2+8 r^3}\right)^2\, ,
\label{K-2}
\end{eqnarray}

In general, the Ricci scalar presents a finite minimum at $r = 0$ and reaches its finite maximum value at another point within the event horizon. In particular when $M \rightarrow \infty$
\begin{eqnarray}
R \rightarrow \frac{180 q^{12}+780 q^{10} r^2+1152 q^8 r^4+768 q^6 r^6+192 q^4 r^8}{q^{14}+4 q^{12} r^2+6 q^{10} r^4+4 q^8 r^6+q^6 r^8}
\, ,
\label{R-2-lim}
\end{eqnarray}
which presents a global maximum at $r = 0$ whose value is $180/q^2$ as well as a finite global maximum value. If also $q \rightarrow \infty$ then $R \rightarrow 0$. For its part, the Kretschmann scalar presents a local maximum at $r = 0$ with a value of $5400/q^4$ in addition to a global maximum at a point located within the interval $0 < r < 5 M/672$. In particular when $M \rightarrow \infty$ we obtain
\begin{eqnarray}
K \rightarrow \frac{8 \left(675 q^{16}+5850 q^{14} r^2+21346 q^{12} r^4+43178 q^{10} r^6+54055 q^8 r^8+43104 q^6 r^{10}+21504 q^4 r^{12}+6144 q^2 r^{14}+768 r^{16}\right)}{q^4 \left(q^2+r^2\right)^8}
\, 
\label{K-2-lim}
\end{eqnarray}
which presents a finite global maximum. Also if  $q \rightarrow \infty$ the $K \rightarrow 0$.


\subsubsection*{$\bullet$ Energy conditions}

Now let us focus on the energy conditions that our second solution satisfies. The WEC is satisfied everywhere for $M > 0$ when $ 0 \leq q \leq 1.2669 M $, given that
\begin{equation}
\rho = \frac{q^2 r }{4 \pi}\left(\frac{48 M^2}{\left(M q^2+8 r^3\right)^2}+\frac{r^2-3 q^2}{\left(q^2+r^2\right)^3}\right) \geq 0
\, ,
\label{wec-s2-1}
\end{equation}
for all $r$ when $0 \leq q \leq 1.4528 M $ and furthermore
\begin{equation}
\rho + p_2 = \frac{2 q^2 r}{4 \pi}\left[- q^2 \left(\frac{72 M^3}{\left(M q^2+8 r^3\right)^3} + \frac{7}{\left(q^2+r^2\right)^3}\right)+\frac{72 M^2}{\left(M q^2+8 r^3\right)^2}+\frac{1}{\left(q^2+r^2\right)^2}+\frac{6 q^4}{\left(q^2+r^2\right)^4}\right]
\, ,
\label{wec-s2-2}
\end{equation}
remains non-negative across its entire domain only when $0 \leq q \leq 1.2669M$. However, the expressions~(\ref{wec-s2-1}) and~(\ref{wec-s2-2}) are positive in the event horizon for $0 \leq q \leq q_{ext}$.

In addition to the previous inequalities, the DEC requires that the following expression remains non-negative
\begin{eqnarray}
&&\rho - p_2 =  \rho - p_3
= \frac{6 q^2 r}{4 \pi} \left[q^2 \left(\frac{24 M^3}{\left(M q^2+8 r^3\right)^3}+\frac{1}{\left(q^2+r^2\right)^3}\right)-\frac{8 M^2}{\left(M q^2+8 r^3\right)^2}-\frac{2 q^4}{\left(q^2+r^2\right)^4}\right]
\, .
\label{dec-s2-3}
\end{eqnarray}
The fact that 
\begin{eqnarray}
\rho - p_2 \rightarrow \frac{6 q^2 }{4 \pi r^5}\left(q^2-\frac{M^2}{8}\right) + O\left(r^{-6}\right)
\, .
\label{}
\end{eqnarray}
as $r \rightarrow \infty$ implies that this quantity is positive if $q > 0.3536 M$. On the other hand $\rho - p_2 \geq 0$ from inside the black hole to infinity when $0.3885 M \leq q \leq 1.1672 M$. Therefore the DEC is fulfilled on the horizon and at infinity when $0.3885 M \leq q \leq 1.1672 M$.

As far as the SEC is concerned, we find that
\begin{eqnarray}
\rho + p_1 + p_2 + p_3 = \frac{q^2}{4 \pi} \left[- 2 q^2 \left(\frac{72 M^3}{\left(m q^2 + 8 r^3\right)^3} + \frac{5}{\left(q^2 + r^2\right)^3}\right) + \frac{96 M^2}{\left(M q^2 + 8 r^3\right)^2} + \frac{1}{\left(q^2 + r^2\right)^2} + \frac{12 q^4}{\left(q^2 + r^2\right)^4}\right]
\, .
\label{}
\end{eqnarray}
This quantity is greater or equal to zero from the event horizon to infinity if $M > 0$ and $q \leq 1.0265 M$. Then, considering the result obtained from Eq.~(\ref{wec-s2-2}) we can establish that the SEC is only satisfied from the event horizon to infinity if $M > 0$ and $q \leq 1.0265 M$.
Within the event horizon there is always a range around zero in which the SEC does not comply.


\section{Two regular black hole solutions that respect the limiting curvature condition}

From the solutions of charged regular black holes obtained in the previous section we can introduce the following spherically symmetric regular black hole space-times or LCC black hole solutions
\begin{equation}
f_I(r) =  1-\frac{432 M^4 r^2}{432 M^4 \ell^2+\left(6 M r+\ell^2\right)^3}
\, ,
\label{1-sol-lcc}
\end{equation}

\begin{equation}
f_{II}(r) = 1-\frac{2 M}{r} \left(1 -\frac{M \ell^2}{M \ell^2+8 r^3}-\frac{\ell^2 r^3}{2 M\left(\ell^2+r^2\right)^2}\right)
\, .
\label{2-sol-lcc}
\end{equation}
Here $M$ is the mass of the black hole and $\ell > 0$ represents the length scale parameter.

We can use the results already obtained by replacing the charge $q$ with the parameter $\ell$ in the previous section to confirm that these solutions do indeed satisfy the limit curvature condition. Then, for the first case the Ricci and Kretschmann scalars are non-singular, even in the limit $M \rightarrow \infty$ where we obtain, respectively
\begin{equation}
R = \frac{12}{\ell^2} \,\,\,\, \mbox{ and } \,\,\,\, K = \frac{24}{\ell^4}
\, .
\label{R-K-sol-I}
\end{equation}
In the second solution, when $M \rightarrow \infty$ we also find that the Ricci and Kretschmann scalars are non-singular in the entire space. In particular, if $\ell$ is very small, the following result is reached, which does not occur at $r = 0$, but does occur in its surroundings
\begin{equation}
R = \frac{192}{\ell^2} \,\,\,\, \mbox{ and } \,\,\,\, K = \frac{6144}{\ell^4}
\, .
\label{R-K-sol-II}
\end{equation}

Regarding the energy conditions satisfied by the corresponding energy-momentum tensors, we can also take the results obtained in the previous section. Consequently, we can assert that the first LCC solution satisfies the WEC (and the NEC) everywhere for all $M>0$ and $\ell > 0$ as observed from Eqs.~(\ref{wec-s1-1}) and~(\ref{wec-s1-2}). Moreover, the DEC holds for all $M>0$ and $\ell > 0$ in the asymptotically flat region. However, the DEC is satisfied in the event horizon only in the narrow range $0.642245 M \leq \ell \leq 0.6458 M$. For its part, the SEC is satisfied in an interval that includes the event horizon and that reaches infinity for $M > 0$ and $\ell \leq 0.6458 M$, that is, in the same range that allows the existence of an event horizon for the black hole solution.

The second LCC solution satisfies the WEC and the NEC for $M > 0$ and $0 < \ell \leq 1.2669 M$. Additionally, the DEC is satisfied in the interval that goes from the event horizon to infinity when $0.3885 M \leq \ell \leq 1.1672 M$. The SEC compliance is possible from the event horizon to infinity when $M > 0$ and $\ell \leq 1.0265 M$.


\section{Circular photon orbits: Quasinormal modes and shadows}

Here we shall consider the propagation of massless particles in a fixed gravitational background, and we shall briefly describe how to compute the radius of the BH shadow as well as the QN spectrum in the eikonal limit of the WKB method.

\subsection{Null geodesics} 

Let us consider the motion of test particles in a given fixed gravitational background characterized by spherical symmetry
\begin{equation}
ds^2 = -f(r) dt^2 + f(r)^{-1} dr^2 + r^2 d \Omega_2^2
\end{equation}
in Schwarzschild-like coordinates ($t,r,\theta,\phi$), where the metric function, $f(r)$, is a known function of the radial coordinate $r$. The geodesic equations are given by \cite{Carroll}
\begin{equation}
\frac{d^2 x^\mu}{ds^2} + \Gamma_{\rho \sigma}^{\mu} \frac{d x^\rho}{ds} \frac{x^\sigma}{ds} = 0,
\end{equation}
with $s$ being the affine parameter, while the Christoffel symbols, $\Gamma_{\rho \sigma}^{\mu}$, are computed by \cite{Carroll}
\begin{equation}
\Gamma_{\rho \sigma}^{\mu} = \frac{1}{2} g^{\mu \lambda} \left( \frac{\partial g_{\lambda \rho}}{\partial x^\sigma}+\frac{\partial g_{\sigma \lambda}}{\partial x^\rho}-\frac{\partial g_{\rho \sigma}}{\partial x^\lambda} \right).
\end{equation}

Since the energy, $E$, and the angular momentum, $L$, defined by
\begin{equation}
E =  f(r) \frac{dt}{ds}, \; \; \; \; \; L = r^2 \frac{d \phi}{ds}
\end{equation}
are constants of motion, the only non-trivial equation is the radial one
\begin{equation}
\left( \frac{dr}{ds} \right)^2 = E^2 - V_{eff}(r)^2, 
\end{equation}
where the effective potential for massless particles (such as photons) is found to be
\begin{equation}
V_{eff}(r)^2 = f(r) \frac{L^2}{r^2}.
\end{equation}
The circular photon orbit, $r_{ph}$, corresponds to the extreme point of the effective potential (see e.g. \cite{Qiao:2022jlu})
\begin{equation}
0 = \frac{d V_{eff}}{dr}|_{r=r_{ph}},
\end{equation}
which is equivalent to the following algebraic equation
\begin{equation}
2 f(r_{ph}) - r_{ph} f'(r)|_{r_{ph}} = 0.
\end{equation}
Then the radius of the BH shadow, $R_{sh}$, as seen by a distant observer, is given by the approximate expression (see e.g. \cite{Konoplya:2020bxa})
\begin{equation}
R_{sh} \approx \frac{r_{ph}}{\sqrt{f(r_{ph})}}.
\end{equation}

\subsection{WKB method in the eikonal limit} 

The QN frequencies, $\omega$, are computed solving the eigenvalue problem of a Schr{\"o}dinger-like equation of the form \cite{valeria,roman,Gonzalez:2022ote,Gonzalez:2021vwp}
\begin{equation}
\frac{\mathrm{d}^{2}\psi(r)}{\mathrm{d}r_{*}^{2}} + \left[\omega^{2} - V(r_*)\right]\psi(r_*) = 0 \ ,
\end{equation}
where the tortoise coordinate is defined by $r_{*}  \equiv  \int \frac{\mathrm{d}r}{f(r)}$, while the potential barrier, $V(r)$, for massless scalar perturbations is computed to be \cite{GBFs}
\begin{equation}
V(r) = f(r)
\Bigg[ 
\frac{l(l + 1)}{r^{2}} + \frac{f'(r)}{r}
\Bigg] ,
\label{poten}
\end{equation}
with $l=0,1,2,...$ being the angular degree, and the prime denotes differentiation with respect to $r$. Finally, the wave equation must be supplemented by the following boundary conditions \cite{valeria,roman}
\begin{equation}
\psi \rightarrow \: \exp( i \omega r_*), \; \; \; \; \; \; r_* \rightarrow - \infty \ ,
\end{equation}
\begin{equation}
\psi \rightarrow \: \exp(-i \omega r_*), \; \; \; \; \; \; r_* \rightarrow  \infty \ .
\end{equation}

In the eikonal regime (i.e., $l \gg 1$) the WKB approximation \cite{wkb1,wkb2} becomes increasingly accurate. Thus, one may obtain analytic expressions for the quasinormal frequencies. In such a limit ($ l \rightarrow \infty$), the angular momentum term is the dominant one in the expression for the potential barrier
\begin{equation}
V(r) \approx \frac{f(r) l^2}{r^2} \equiv l^2 g(r) ,
\end{equation}
where for convenience we have defined a new function $g(r) \equiv f(r)/r^2$. 
%
%
The maximum of the potential, ($r_1,V_0$), is obtained finding the root of the following algebraic equation
\begin{equation}
2 f(r_1) - r_1 f'(r)|_{r_1} = 0,
\end{equation}
and therefore $r_1=r_{ph}$.
The idea and formalism were treated in \cite{eikonal1}. The QNMs, in the eikonal regime, are found to be
\begin{equation}
\omega(l \gg 1) = \Omega_c l - i \left(n+\frac{1}{2}\right) |\lambda_L| ,
\end{equation}
where $n=0,1,2,...$ is the overtone number, and the Lyapunov exponent $\lambda_L$ is given by \cite{eikonal1}
\begin{equation}
\lambda_L = r_1^2 \sqrt{\frac{g''(r_1) g(r_1)}{2}},
\end{equation}
while the angular velocity $\Omega_c$ at the unstable null geodesic is given by \cite{eikonal1}
\begin{equation}
\Omega_c = \frac{\sqrt{f(r_1)}}{r_1} = \frac{1}{R_{sh}}.
\end{equation}
We comment in passing that applying the WKB approximation of 1st order \cite{MassiveG}
\begin{equation}
\frac{i Q(r_1)}{\sqrt{2 Q''(r_1)}} = n + \frac{1}{2},
\end{equation}
where by definition $Q(r) = \omega^2 - V(r)$, one may obtain the same expression reported before for 
$\{ \Omega_c, \lambda_L \}$, see for instance \cite{Konoplya:2020bxa}
\begin{equation}
\omega_n^2 = V_0 - \sqrt{-2 V''_0} \: \left( n+\frac{1}{2} \right) i,
\end{equation}
Our numerical results are summarized in Table 1 for both solutions discussed here. We have assumed $M=1$, and two different values of the electric charge, $q_1=0.6423,q_2=0.6458$, within the allowed range so that all conditions are satisfied. We observe that regarding solution 1 a higher electric charge implies a higher angular velocity (and consequently a lower radius of BH shadow), and a lower Laypunov exponent. Regarding solution 2 however, a higher electric charge implies a higher Lyapunov exponent. Moreover, for a given electric charge, the second solution is characterized by a lower angular velocity, and a higher Lyapunov exponent and radius of BH shadow.


\begin{table}
\centering
	\caption{Angular velocity $\Omega_c$, Lyapunov exponent $|\lambda_L|$ and radius of BH shadow $R_{sh}$ considering $q_1=0.6423$, $q_2=0.6458$ and $M=1$ for the two LCC solutions discussed here. The first and third lines correspond to the first solution, while the second and fourth lines to the second solution.}
	\label{tab:1}       
	\begin{tabular}{cccc}
		\hline\noalign{\smallskip}
		Electric charge $q$  & $\Omega_c$ & $|\lambda_L|$ & $R_{sh}$ \\
		\noalign{\smallskip}\hline\noalign{\smallskip}
		0.6423  &  0.2182  &  0.1677  & 4.5833  \\ 
		0.6423  &  0.2069  &  0.1983  & 4.8339  \\ 
		0.6458  &  0.2186  &  0.1669  & 4.5748   \\
		0.6458  &  0.2070  &  0.1984  & 4.8301   \\
	\noalign{\smallskip}\hline
	\end{tabular}
\end{table}


\section{Conclusions}

To summarize our work, in the present article, we have obtained for the first time regular charged black hole solutions in four dimensions that satisfy the Limiting Curvature Condition \cite{Markov:1982, Markov:1984ii, Polchinski:1989ae}. According to this conjecture, the spacetime curvature should always be restricted by some universal value ($|R| \leq \mathcal{B}\ell^{-2}$, where $R$ is a scalar curvature invariant of dimension [length]$^{-2}$, 
$\mathcal{B}$ is a dimensionless constant that may depend on the type of curvature invariant (but remains independent of specific solutions within the theory), and the parameter $\ell$, which is related to the radius of curvature of a fundamental length of the theory). 

\smallskip

Based on the aforementioned conjecture, we have considered two models described by General Relativity coupled to the appropriate non-linear electrodynamics (as can be verified in Eqs. \ref{Lagr-1} and \ref{Lagr-2}), and we have obtained two new regular charged black hole solutions. For both solutions we have studied the dynamic stability, the scalar invariants $R, K$, and the usual four energy conditions. We then confirm by direct inspection that both cases satisfy the limit curvature condition (i.e. in both cases we obtain $R \propto \ell^{-2}$ and $K \propto \ell^{-4}$). Finally, in order to see their differencies in astrophysical implications, we have given a brief summary of null geodesics, circular photon orbits and radius of BH shadow as well as the QN spectrum within the WKB approach in the eikonal limit. Our numerical results were summarized in Table 1.

\smallskip

We have assumed a fixed BH mass $M=1$, and two different values of the electric charge, $q_1=0.6423,q_2=0.6458$, within the allowed range so that all conditions are satisfied. Our findings indicate that regarding solution 1 a higher electric charge implies a higher angular velocity (and consequently a lower radius of BH shadow), and a lower Laypunov exponent. Regarding solution 2, however, a higher electric charge implies a higher Lyapunov exponent. Moreover, for a given electric charge, the second solution is characterized by a lower angular velocity, and a higher Lyapunov exponent and radius of BH shadow.

\section*{Acknowledgements}

We wish to thank the anonymous reviewers for their useful comments and suggestions.
A.~R. acknowledges financial support from Conselleria d'Educació, Cultura, Universitats i Ocupació de la Generalitat Valenciana through Prometeo Project CIPROM/2022/13.
A.~R. is funded by the Mar{\'i}a Zambrano contract ZAMBRANO 21-25 (Spain).
The author L.~B. is supported by DIUFRO through the project: DI24-0087.


\end{document}